# Un enfoque de enseñanza de la Astronomía:
## Algunas consideraciones epistemológicas y didácticas

Alejandro Gangui y Agustín Adúriz-Bravo

## Introducción

Para orientar la tarea docente en Inicial y Primaria, nos parece fundamental partir de un análisis epistemológico y didáctico de los contenidos programáticos del Área del Conocimiento de la Naturaleza en el currículo actualmente vigente en Uruguay. En este artículo realizamos una breve discusión de los contenidos de Astronomía prescritos en el *Programa de Educación Inicial y Primaria: Año 2008* (ANEP, 2013), con el objeto de brindar a las maestras y maestros algunas "claves de lectura" que se traduzcan en posibles directrices para el diseño, implementación y evaluación de secuencias de enseñanza que aborden esos contenidos.

En todo diseño curricular se pone en juego una gran variedad de criterios de selección de contenidos: se espera que ellos sean centrales en la(s) disciplina(s) de referencia, socioculturalmente significativos, adecuados al desarrollo psicoevolutivo de los estudiantes, valiosos para la formación de ciudadanía, etc. Entre tales criterios necesariamente se generan ciertas tensiones. El conjunto de contenidos seleccionados para ser enseñados, el orden en el que aparecen en los distintos grados, la importancia relativa que se les asigna, entre otros muchos aspectos, dan cuenta de la resolución que se ha dado a esas tensiones: el currículo no solo muestra la imagen de ciencia que se desea "hacer vivir" en las aulas, sino que está "teñido" de las concepciones epistemológicas y didácticas sostenidas por sus diseñadores. Examinar el currículo como el "producto final" de delicadas decisiones y negociaciones nos proporciona una "ventana" a entender tanto qué se ha considerado importante enseñar como en qué se fundamenta el reconocimiento de esa importancia.

El proceso de producción del currículo como documento público conlleva poner en diálogo, y eventualmente lograr amalgamar, los supuestos e intenciones asumidos por los diseñadores, que provienen de diferentes campos profesionales, tradiciones intelectuales y disciplinas académicas. Leer con cuidado los contenidos que se prescriben nos permite indagar en ese proceso para fundamentar mejor la enseñanza, y tal es la intención de este artículo.

## La "epistemología" debajo del programa de Astronomía

Al analizar el brevísimo programa en su conjunto, lo primero que se constata es que en el documento se "trenzan" contenidos provenientes de diferentes tradiciones disciplinares, que a lo largo de la historia de la ciencia han estado más cercanas o más alejadas unas de las otras (e incluso a veces en conflicto).[1] Estas tradiciones se pueden reconocer como "hilos conductores" del programa; cada uno de esos hilos arranca en un determinado grado de escolaridad –más o menos tempranamente– y luego se sostiene a lo largo de los demás, ampliándose y profundizándose.

En primer lugar, en el programa figura centralmente la Astronomía más tradicional, descriptiva y matemática, que es probablemente la primera de las ciencias naturales en emerger, y que en la Antigüedad estuvo ligada principalmente a tres grandes intereses humanos: la medición y el control de los ciclos naturales del tiempo (años, estaciones, días), las (supuestas) relaciones del mundo celeste con el terrestre (predicciones astrales), y la navegación en sentido amplio (orientación en tierra o en el mar). En esta Astronomía en sentido más "limitado" se ubicarían los contenidos protagónicos del programa, vinculados con los movimientos de la Tierra: rotación (alternancia día-noche y relojes) y traslación (estaciones y calendarios). Tales contenidos son muy abundantes.

El programa prescribe trabajar este primer eje astronómico poniendo el foco en el "sistema" Tierra-Sol: la relación entre ambos cuerpos, principalmente en términos de sus posiciones relativas. También propone observar y describir con cuidado muchos aspectos de este primer eje: cielo diurno y nocturno, sombras, duración del día y la noche, estaciones, y cambios de todo ello a lo largo del día o del año. Se indica, por ejemplo, "seguir" el lugar de salida y puesta del Sol y ver sus variaciones periódicas. También hay trabajo sugerido en torno a la observación y descripción de la bóveda celeste, y en especial del cielo nocturno (incluyendo algunas constelaciones icónicas visibles desde nuestro hemisferio).

A este primer eje se suman temas que podríamos ubicar en la interfaz de la Astronomía con otras ciencias naturales (especialmente con la física), tal es el caso de los contenidos de mecánica, óptica, astrofísica y cosmología. Así, se prescribe el estudio de: ciertas interacciones entre Sol, Tierra y Luna (fases, eclipses, mareas); la identificación y descripción de algunos "tipos" de objetos celestes (galaxias, estrellas, planetas, satélites naturales), cruzando esto con la noción de radiación electromagnética; la implantación de una mirada "evolutiva" (origen y decurso del Sistema Solar y del Universo).

También aparecen, aunque en menor medida, contenidos conceptuales más propios de la física o la meteorología que tendrían un carácter "instrumental" para el pensamiento astronómico: radiación proveniente del Sol, brillo y color estelar, diferencias térmicas en el día y en el año.

Por otra parte, algunos contenidos son de carácter procedimental, y nos dan una idea del tipo de "didáctica" que imaginaron los diseñadores. Aquí cabrían la identificación de las "zonas del horizonte" y los puntos cardinales, el aprendizaje de algunas estrategias de orientación con el

---

[1] La historia de la enseñanza de la Astronomía en el Uruguay no es objeto del presente artículo, pero es sumamente interesante y puede encontrarse en Pintos Ganón y Fernández (2008).

Sol y las estrellas, o la familiarización con distintas representaciones de la Tierra (probablemente cerca de la geografía y la cartografía).

Y por último, también hay algunos contenidos del campo de la historia y la naturaleza de la ciencia (Adúriz-Bravo, 2005): en quinto grado se incluye la discusión entre los modelos geocéntrico y heliocéntrico y en sexto aparecen las "teorías" sobre el origen y la evolución del Universo. Aunque la formulación es poco clara, entendemos que allí se sugiere a las maestras y maestros no solo trabajar la *cosmogonía*, es decir, el conjunto de relatos mitopoéticos o religiosos que las distintas civilizaciones han elaborado para explicar el origen y naturaleza del "mundo" (Gangui, 2016), sino además ofrecer una primera aproximación a las modernas teorías sobre la verdadera constitución física del cosmos (Gangui, 2005).

Con todo lo que hemos venido diciendo, cabe reconocer que la Astronomía como disciplina escolar tiene rasgos constitutivos muy propios (diferentes modos de explicación, distintas perspectivas teóricas, multitud de ciencias auxiliares, diversidad de escalas de espacio y de tiempo); tener presente esta peculiaridad es de gran valor para pensar los fundamentos de la enseñanza. A ello se suma la particular naturaleza epistemológica de la Astronomía como una ciencia en la que, a las explicaciones mecánicas y causales más "ortodoxas", se suman explicaciones funcionales (en torno a estructura, funcionamiento e interrelación) y genéticas (en torno a origen y evolución). Los rasgos de la asignatura escolar, parcialmente heredados de la disciplina académica, se pueden entender tratando de caracterizar la diversidad de preguntas que se pretende contestar desde la "mirada astronómica" en el nivel inicial y primario:

1) ¿Dónde estamos? ¿Qué/cómo son la Tierra, el Sistema Solar y el Universo? ¿Cómo podemos orientarnos y viajar en ellos?
2) ¿Cuál es el origen de esos tres sistemas anteriores? ¿Cuál ha sido y será su evolución a lo largo del tiempo? ¿Y cuál será su final?
3) ¿Qué ejemplos relevantes hay de interacciones (de todo tipo) entre las partes que constituyen esos sistemas?
4) ¿Cómo podemos describir y representar esos sistemas de manera estática y dinámica? ¿Qué "hechos del mundo" nos permiten conocer, explicar y predecir tales descripciones y representaciones?
5) ¿Qué diferentes visiones ha generado la humanidad sobre esos sistemas? ¿Cómo se han modificado tales visiones a lo largo del tiempo? ¿Cuáles visiones están actualmente vigentes?

**Algunos aportes desde la didáctica de la Astronomía**

¿Qué contenidos enseñamos cuando enseñamos Astronomía? Como docentes, formulamos esta pregunta cada vez que damos inicio a un curso o actividad sobre educación en temas astronómicos. Usualmente, las respuestas de los participantes incluyen tópicos variados: el Sistema Solar, las galaxias, estrellas y constelaciones, el *Big Bang*, la forma y los movimientos de la Tierra, las órbitas planetarias, y la lista continúa. En función de estas respuestas notamos que, ante las preguntas sobre una disciplina científica, los docentes tendemos a desplegar –al menos en un primer momento– una lista de términos y conceptos, como si eso fuese lo único importante a considerar en la enseñanza. Así, muchas veces relegamos otros aspectos esenciales del conocimiento científico, más vinculados con la dimensión metodológica de la ciencia: por ejemplo, la formulación de preguntas, el planteamiento de hipótesis, la discusión de ideas, la exploración sistemática de fenómenos, la construcción colectiva del conocimiento, entre otros. Los conceptos son importantes, nadie lo duda, ¿pero acaso son suficientes como para que los estudiantes se acerquen al conocimiento escolar? ¿Y en qué casos esto sería así?

Para comenzar a reflexionar sobre estas cuestiones, pensemos en la siguiente situación que ocurrió hace un poco más de cuatro años en una escuela pública de la ciudad donde vivimos los autores de este artículo. Como parte de su trabajo diario en las escuelas, una colega nuestra, la profesora María Iglesias, participó de una clase de ciencias naturales en la que estudiantes de sexto grado de la escuela primaria argentina (de 11-12 años de edad) realizaban un "intercambio de cierre" sobre uno de los temas estudiados con compañeros del tercer grado de la misma escuela. El núcleo temático que motivó esta actividad fue la Astronomía, ya que ambos cursos tuvieron una aproximación a contenidos de esa área. Durante el recorrido didáctico previo, uno de los grupos había estado abocado al trabajo con las propiedades ópticas de los objetos, la producción de sombras y el movimiento aparente del Sol durante un día solar. El grupo de niños más grandes, por su parte, se había dedicado a profundizar en el recorrido aparente del Sol a lo largo del año, para diferentes ubicaciones del observador terrestre, ya fuera en el hemisferio sur o en el norte (temas estos que aparecen en el programa uruguayo).

En este intercambio se alentaba a las niñas y niños para que expresaran lo aprendido durante las actividades transitadas, en las que fueron guiados por los docentes de cada grado. A partir de los diálogos entre los estudiantes, quedaba claro que el objetivo de los docentes durante la enseñanza de los contenidos había estado centrado en los aspectos más vivenciales y descriptivos de los fenómenos astronómicos, en contraposición a los elementos "expositivos" que se encuentran de manera más habitual en la enseñanza del área. Para dar inicio a los recorridos didácticos implementados se habían propuesto actividades para conocer y hacer explícitos los saberes previos sobre el tema. Los docentes hallaron que la mayoría de los estudiantes respondió (¿o repitió una vez más?) sin poner en duda que es la Tierra la que se mueve alrededor del Sol.

Pensemos un poco: ¿los estudiantes estaban en condiciones de usar ese conocimiento inicial sobre el movimiento de la Tierra para analizar, por ejemplo, a qué se debe que el Sol del mediodía, a lo largo del año, no tenga siempre la misma altura por encima del horizonte del observador? Decir que la Tierra rota sobre su eje y que se traslada, ¿es suficiente para inferir que, para nuestras latitudes fuera de la banda de los trópicos –y esto incluye todo el territorio de Uruguay–, el Sol nunca "pasa" por encima de nuestras cabezas?

Si tenemos en cuenta las respuestas ofrecidas por los estudiantes durante el desarrollo de la secuencia de enseñanza en relación con el movimiento aparente del Sol, podemos decir que no estaban en condiciones de emplear adecuadamente ese conocimiento, de carácter *declarativo* (Aparicio, 1995). En particular, en el trabajo se encontraron evidencias del desconocimiento de la trayectoria aparente del astro, de las regularidades de su movimiento y de los cambios que presenta con el correr de los días. Muchos de los niños y niñas, además, tenían dificultades para establecer relaciones entre el movimiento del Sol y la producción de las sombras de objetos opacos.

Como podemos imaginar, las ideas anteriores, relacionadas con el sistema Sol-Tierra y prescritas en el programa que estamos analizando, apuntan a un nivel de abstracción difícil de lograr con niños de estas edades (hasta 12 años). Sin embargo, parece claro que la propuesta de trabajar *a partir de las vivencias astronómicas del observador terrestre* resulta mucho más rica que el mero hecho de declarar cuáles son los movimientos reales que tiene nuestro planeta. Sobre estas cuestiones, Martínez Sebastià (2004) plantea que la enseñanza de la Astronomía no puede limitarse a saber que la Tierra es una esfera que gira sobre sí misma y que se traslada alrededor del Sol. Su objetivo, en cambio, es que los estudiantes aprendan de manera progresiva a explicar hechos y fenómenos astronómicos del entorno mediante esas dos nociones entendidas como genuinas *hipótesis*. Entonces, cuando escuchamos a nuestros estudiantes afirmar que es la Tierra la que se mueve –y no el Sol, por ejemplo–, debemos cuestionarnos si no se trata más de una *idea adquirida* que de una *idea* genuinamente *construida*.

**Ideas previas y obstáculos de comprensión en la Astronomía escolar**

¿Cuáles son los modelos teóricos y los tipos de explicación más frecuentes que utilizan nuestros estudiantes de nivel inicial y primario en relación con temas de Astronomía? ¿De qué manera estas ideas afectan la construcción del conocimiento científico? Numerosas investigaciones realizadas en varios países ponen de manifiesto los obstáculos que se presentan en relación a algunos de estos fenómenos (Gangui e Iglesias, 2015). Entre los más frecuentes podemos mencionar las dificultades de las niñas y niños al intentar reconocer los cambios en los aspectos observables del movimiento del Sol, tanto a lo largo del día como del año. Solo una baja proporción de estudiantes sabe que la trayectoria diaria del Sol tiene forma de arco con extremos en el horizonte. Por otra parte, muy pocos son capaces de identificar las características particulares de los solsticios y los equinoccios, y las regularidades en torno a ellos, lo cual va asociado a una visión distorsionada de cómo se modifica el cielo observable con el correr de las semanas. Asimismo, los estudiantes en general presentan dificultades para reconocer la existencia de distintos modelos alternativos *válidos* que pueden dar cuenta de las mismas observaciones (cambiando el punto de referencia, por ejemplo) y no hacen uso *operativo* de las hipótesis de esos modelos para explicar las observaciones conocidas.

Las explicaciones astronómicas sobre la Luna, con sus fases, y el movimiento espacial de los tres cuerpos celestes involucrados (Sol-Tierra-Luna) resultan el tópico en torno al cual se reportan mayores dificultades. Como sabemos, las fases de la Luna que percibimos desde nuestro lugar en la Tierra son vivencias astronómicas *topocéntricas.*[2] Sin embargo, para dar una explicación sobre su origen es frecuente recurrir a un punto de vista situado en el espacio interplanetario. Por lo tanto, resulta imprescindible imaginarse cómo se verían las fases de la Luna desde un punto de observación ubicado fuera de la Tierra. Allí reside el problema, pues uno de los principales obstáculos de comprensión de la Astronomía es el que está ligado a la visión espacial, es decir, a la capacidad mental de ver y trabajar en tres dimensiones y con tamaños y distancias enormes. Además, las distancias que separan al Sol y a la Luna de nosotros, los observadores, hacen imposible su medición directa desde la Tierra, y esto lleva a que los estudiantes emitan juicios acerca de la lejanía de esos astros sobre la base del tamaño relativo que ellos perciben (García Lerete, 2010).

A estas dificultades debemos sumar las ideas previas que han sido estudiadas *in extenso* por la literatura específica. Como es sabido, las personas poseen una tendencia natural a investigar y a realizar acciones en el medio en el que viven, lo que las conduce a construir explicaciones espontáneas para dar sentido a los fenómenos observados. A lo largo de los años, el conocimiento intuitivo, vale decir, las explicaciones o ideas que construimos los individuos, han recibido diferentes denominaciones (ideas previas, concepciones alternativas, ideas ingenuas, teorías implícitas, modelos mentales, etc.). De manera general, podemos considerar todo aquello como las "concepciones" que tienen las personas –en particular, los estudiantes– acerca de cómo y por qué las cosas son como son. Las concepciones responden a una lógica "interna" de pensamiento, influenciada por las experiencias realizadas en la vida cotidiana; tenemos entonces ideas de origen sensorial –*concepciones espontáneas*–, cultural –*representaciones culturales*– o escolar –*concepciones analógicas*–. Todas estas ideas forman parte de lo que el sujeto ya sabe, y pueden generar interferencias que dificultan el proceso de construcción científica. Este conocimiento –y los supuestos en los que se basa– puede reestructurarse, pero rara vez se abandona, en tanto que, como señala Ignacio Pozo (1999), presenta gran eficacia cognitiva y adaptativa para "movernos por el mundo".

Entre los muchos fenómenos astronómicos que presentan dificultades para su comprensión significativa podemos mencionar los siguientes:

1) Las fases de la Luna: las diferentes iluminaciones de la superficie de nuestro satélite, ¿son debidas a la sombra de la Tierra? Estas fases, ¿surgen como consecuencia de que, desde el punto de vista de la Luna, se produce un eclipse, donde es la Tierra el cuerpo astronómico que oculta al Sol?

---

[2] "Topocéntrico" se refiere a la posición precisa *sobre la superficie de la Tierra* desde donde se lleva a cabo la observación, en contraposición con, por ejemplo, "geocéntrico", que se refiere a la descripción efectuada desde el centro de nuestro planeta. Recordemos que *topos* en griego significa "lugar" (en este caso, ubicación sobre la superficie terrestre).

2) El ciclo día-noche: el eterno ciclo de luz y sombra, de días y noches, ¿se debe al movimiento de la Tierra alrededor del Sol? ¿Dónde va el Sol al anochecer? ¿Se esconde detrás de las montañas o de las nubes?

3) La verticalidad en la Tierra y la gravitación; en particular, el problema de los *antípodas*[3] y la concepción de la gravitación como atracción hacia el centro del planeta: ¿qué significa caer "hacia abajo"? ¿Dónde queda ese "abajo"?

4) Las cuatro estaciones: las diferentes estaciones del año, con sus climas y temperaturas característicos, ¿se producen debido a que la Tierra se halla a diferentes distancias del Sol? ¿O es que, para una ubicación geográfica dada, el eje de la Tierra se inclina más en verano que en invierno?

5) La composición y forma del Sistema Solar: las trayectorias planetarias son usualmente representadas con formas pronunciadamente elongadas, lo que hace que se vean como elipses de notable *excentricidad*[4] (y aquí existe una relación con las diversas nociones sobre las temperaturas en las diferentes estaciones del año). Además, ¿dónde termina nuestro Sistema Solar, en la próxima estrella?

6) Nuestra ubicación en el Universo, es decir, la posición de la Tierra, la del Sistema Solar y la de nuestra galaxia: ¿ubicación?, ¿respecto de qué? ¿Existen un arriba y un abajo en el Universo?

Las ideas previas en torno a todos estos temas están presentes en cualquier situación de aprendizaje: influyen en las observaciones que realizamos, en las inferencias que construimos y en la manera de resolver problemas. En las investigaciones se señala que la recurrencia de tales ideas se prolonga en la escuela secundaria y más allá, afectando también a estudiantes de magisterio y a profesores en actividad (Camino, 1995; Gangui e Iglesias, 2015).

**A modo de síntesis**

En el *Programa de Educación Inicial y Primaria* se exponen los fundamentos legalmente prescritos para abordar la enseñanza de la Astronomía en los primeros nueve niveles del Sistema Educativo. A través de un rápido análisis de su diseño, hemos podido identificar algunas ideas-fuerza que tienen gran potencia didáctica: la Astronomía escolar reúne distintas tradiciones disciplinares, busca responder preguntas teóricas muy disímiles, se enfoca en sistemas que tienen diferentes escalas de espacio y tiempo, responde a finalidades humanas (históricas y actuales) bien reconocibles.

---

[3]Literalmente, *antípoda* significa "opuesto por los pies". Si trazamos una recta que, partiendo de nuestra ubicación sobre la superficie de la Tierra, atraviese su centro, el lugar por donde vuelve a tocar la superficie corresponde a nuestros antípodas. Los antípodas de los habitantes de la ciudad de Montevideo nadan en medio del mar Amarillo, hacia el oeste de Corea del Sur.

[4]La excentricidad de la órbita (designada con la letra *e*) indica cuánto se aparta esta órbita de una circunferencia perfecta (que tiene e=0). Un valor de 0<e<1 corresponde a una órbita elíptica.

A lo largo de este artículo también hemos presentado, de manera sucinta, algunos elementos básicos relacionados con los modelos teóricos y los tipos de explicación más frecuentes que utilizan los estudiantes en relación con los temas más centrales de la Astronomía escolar.